\documentstyle[12pt]{article}
\input epsf
\newcommand{\lash}[1]{\not\! #1 \,}

\begin{document}
\renewcommand{\thefootnote}{\fnsymbol{footnote}}
\begin{titlepage}
\renewcommand{\thefootnote}{\fnsymbol{footnote}}
\makebox[2cm]{}\\[-1in]
\begin{flushright}
\begin{tabular}{l}
TUM/T39-96-10
\end{tabular}
\end{flushright}
\vskip0.4cm
\begin{center}
{\Large\bf
Physically Motivated Approximation to a Parton\\
Distribution Function in QCD\footnote{Work supported in part by BMBF}} 

\vspace{2cm}

L.\ Mankiewicz\footnote{On leave of absence from N. Copernicus
Astronomical Center, Polish Academy of Science, ul. Bartycka 18,
PL--00-716 Warsaw (Poland)} and T. Weigl 

\vspace{1.5cm}

{\em Institut f\"ur Theoretische Physik, TU M\"unchen, Germany}

\vspace{1cm}
 
{\em \today}
 
\vspace{1cm}
 
{\bf Abstract:\\[5pt]}
\parbox[t]{\textwidth}{
It has been suggested that parton distributions in coordinate space, so
called Ioffe-time distributions, provide a more natural object for
non-perturbative methods compared to the usual momentum distributions.
In this paper we argue that the shape of experimentally determined
Ioffe-time distributions of quarks in a nucleon target clearly indicates
separation of longitudinal scales, which is not easily
recognizable in terms of conventional longitudinal momentum space
considerations. We demonstrate how to use this observation to
determine parton distributions, using
non-perturbative information about the first few moments and the
Regge asymptotics at small $x$.
}
 
\vspace{1cm}
{\em Published in }Phys.Lett. {\bf B380}, 134 (1996) 
\end{center}
\end{titlepage}
 
\newpage

Deep inelastic scattering provides one of the cleanest applications of
perturbative QCD. According to factorization theorems \cite{Col89},
the entire $Q^2$ dependence of the cross section can be calculated
perturbatively, while all dynamical effects of large distances can be
parametrised by a set of one-particle parton distribution functions
given at a certain reference scale. Although these parton distributions
are determined experimentally with a good accuracy, their calculation
from first principles remains a challenge for non-perturbative QCD
methods.

In the past decade a remarkable progress has been made at the experimental
side, and apart from the region of very small Bjorken $x$, there is now not
much controversy regarding the existing parametrizations of parton
distributions \cite{Blum95}. The theoretical progress has been much slower.
Apart from several quark-model or MIT bag model calculations, there have been
relatively few attempts to determine parton distributions from QCD. The problem
has proved to be very difficult for the theory. QCD sum rules calculations of
properties of parton distributions have been moderately successful
\cite{SumRules,Ros96}.  The state-of-art lattice QCD calculation of the lowest
moments of quark distribution functions has appeared just recently
\cite{Lat95,Lat96}. In the present paper we propose an approximate scheme to
compute bulk of momentum space quark distributions starting from relatively
modest theoretical information. The approximation can be systematically
improved when new information becomes available. Note that we are not primarily
concerned here with an effective mathematical method of reconstruction of a
parton distribution from its (many) moments. Such methods exists
\cite{Par79,Fur82,Kat94} and are very useful in the perturbative QCD analysis
of experimental data. In this paper we consider a different problem of
computation of structure function from a non-perturbative theoretical input.
Hence, anticipating technical difficulties in calculation of higher moments
from QCD, we have been looking for a method which can give a satisfactory
description using a minimal amount of theoretical information.

Parton distribution functions arise from long-distance QCD physics which is
still perhaps the least understood domain of strong interactions. Given the
intrinsic complexity of the problem it is clear that an approximation scheme is
necessary. The main problem is to understand what information is required to
compute a parton distribution function in QCD with a reasonable accuracy.
Usually modern parametrizations of parton densities \cite{MRS,CTEQ,GRV94} rely
on input distributions assumed to be valid at some low normalization scale.
Parton distributions at any larger scale can be uniquely determined thanks to
the QCD evolution equations.  As very little is understood at present about the
low scale input distributions, they are usually assumed to follow a simple
shape which is adjusted iteratively to reproduce the experimental data.
Alternatively one can calculate them in a certain QCD motivated model, but the
significance of such calculation is not obvious. Thus, the question we address
in this paper can be formulated as follows: how to determine input
distributions making the best use of the information available from
state-of-art non-perturbative QCD calculations.

It is a textbook statement that twist-2 parton distribution functions describe
probability densities of partons longitudinal momenta. Furthermore it is
usually assumed that the connection between moments of parton distributions and
matrix elements of local twist-2 operators, provided by OPE \cite{Col82,Bal88},
should make it possible to compute them, say, in the lattice QCD approach. Our
experience with QCD sum rules and understanding of the present status of the
lattice QCD technology tells us however that in reality only the lowest moments
are computable with a reasonable accuracy, assuming that available resources
will not suddenly increase by {\em many} orders of magnitude. While we realize
that such assumption has proved wrong many times in the history, we have found
it nevertheless justified to look for a non-standard scheme which results in a
good approximation to the parton densities starting from a very modest amount
of information. As we shall show below, such a scheme can be derived from the
analysis of {\em experimental} properties of parton distributions. We shall
argue that the amount of information relevant to quark distribution functions
is surprisingly small, even though a crucial piece cannot be easily obtained
using the standard QCD methods.

In this context it turns out to be advantageous to analyze longitudinal
distance - or Ioffe-time - distributions rather then more common longitudinal
momentum distributions. Longitudinal distance distributions were introduced
many years ago \cite{LDD}, but until very recently \cite{Duc92,Bra95} their
significance has not been fully understood.  Mathematically Ioffe-time
distributions are just Fourier transformed longitudinal momentum distributions.
Let $q^+(u,\mu^2) = q(u,\mu^2) + {\bar q}(u, \mu^2)$ denote the positive
C-parity quark longitudinal momentum distribution at a scale $\mu^2$.  The
corresponding Ioffe-time distribution can be defined as
\begin{equation}
Q^+(z,\mu^2) = \int_0^1 du\, q^+(u,\mu^2) \sin (uz) \, .
\label{def1}
\end{equation} 
Because in this paper we concentrate on positive C-parity distributions the
superscript $^+$ will be neglected in the following.  As we are primarily
concerned with low-scale distributions, $\mu^2$ is always equal to $4$ GeV$^2$,
unless it is explicitly indicated. Although it is not possible to gain any
information by transforming a momentum distribution into coordinate space, we
shall show that certain phenomenological properties of quark distributions are
easier to grasp in the Ioffe-time representation.  The coordinate space
variable $z$ is the invariant measure of the longitudinal light-cone distance
between the points where the hard probe was absorbed and emitted by the target.
In the leading logarithmic approximation $Q(z)$ has a very simple and
transparent interpretation as being related to the target matrix element of a
non-local QCD string operator \cite{Col82,Bal88}. Indeed, let $\Delta$ denote a
light-like vector, $\Delta^2 =0$. If $P$ denotes a target momentum, then the
Ioffe-time $z = P \cdot \Delta$ and $Q(z,\mu^2)$ can be defined as
\begin{equation}
\langle P \mid {\bar \Psi}(\Delta)\lash{\Delta} [\Delta;0] \Psi(0)
\mid P \rangle_{\mu^2} - (\Delta \to - \Delta)  =   
4 i (P \cdot \Delta)\, Q(z, \mu^2) \, .
\label{QQQ}
\end{equation}
where $[\Delta;0]$ is the path-ordered exponential necessary to insure gauge
invariance.  The evolution equations for Ioffe-time distributions were
considered in \cite{Bal88,Bra95}.  It is important to realize that derivatives
of Ioffe-time distribution $Q(z)$ at the origin are given by moments of
corresponding structure functions $q(u)$, or equivalently \cite{Col82,Bal88} by
matrix elements of twist-2 operators of increasing dimension.  It has been
argued \cite{Bra95} that from the theoretical point of view this connection
makes Ioffe-time distributions much easier to analyze than parton distributions
in momentum space.

Typical shapes of u and d-quark Ioffe-time distributions $Q_u(z)$ and $Q_d(z)$
obtained with the help of existing parton parametrizations
\cite{MRS,CTEQ,GRV94} are shown in figures 1 and 2. As expected different
parametrizations \cite{MRS,GRV94} produce very similar Ioffe-time
distributions, although, somewhat surprisingly, parametrization \cite{CTEQ}
results in different behavior for large $z$. The value of $Q(z)$ at $z =
0$ equals zero by definition if the momentum distribution $q(u)$ is less
singular than $u^{-2}$ at $u = 0$. A much less trivial observation concerns the
shape of $Q(z)$. In the region of small $z$, say up to $z \sim 3-4$, the
distribution is smooth, to a good approximation linear function of $z$.  Note
that its slope at the origin is equal to the longitudinal momentum fraction
carried by quarks. The behavior of $Q(z)$ in this region is determined by {\it
  average} properties of the corresponding longitudinal momentum distribution
which are encoded in its few lowest moments.  A change of the shape of $Q(z)$
in this region results in a change of the bulk properties of the momentum
distribution. Note that $z=10$, i.e. the onset of the asymptotic behavior,
corresponds in the nucleon rest frame to a longitudinal distance of
the order of 2 fm, or the nucleon diameter. Having in mind a simple
geometrical picture one can argue that for larger values of $z$
absorbtion and emission of the virtual photon by the target occurs outside
the space-time volume occupied by the nucleon. For
larger values of $z$ absorption respectively emission of the virtual photon by the
target occurs outside the space-time volume occupied by the nucleon.

In this asymptotic domain $Q(z)$ behaves according to 
{\it small}-$u$ behavior of the longitudinal momentum distribution. Indeed, if
$q(u) \sim u^\alpha$ at small $u$, then from (\ref{def1}) it follows that $Q(z)
\sim 1/z^{1+\alpha}$ for large $z$. A change of shape of $Q(z)$ in this region
would influence the small-$u$ behavior of the momentum distribution.

Physically the behavior at large-$z$ reflects properties of {\em wee} partons,
while the small-$z$ region is sensitive to the distribution of {\em hard}
partons i.e., those which carry finite longitudinal momenta. The presence of a
clearly recognizable transition between these two regions, see figures 1 and 2,
suggests that they should be treated separately. From the mathematical point of
view the Taylor expansion of $Q(z)$ around point $z = 0$ has infinite radius of
convergence, so formally the asymptotic region can be reached from the origin.
It corresponds to the fact that the inverse Mellin transformation always allows
us to reconstruct the momentum distribution from its moments. In practice it
requires to know a lot of them - many more than we can perhaps ever hope to
have computed on the lattice. To illustrate this point on figure 3 we show the
convergence of the short-distance expansion of $Q_u(z)$. 
A line labeled Tn denotes the
Taylor approximation constructed from the first n non-vanishing moments. 
It can be
seen that while T1 is a good approximation to $Q(z)$ in the small-z domain, it
takes a lot of moments to reach the asymptotics. Hence the idea, first proposed
in \cite{Bra95}, to consider each of these two regions separately, and then
match them in the transition region. Lattice QCD is certainly the best approach
to calculate $Q(z)$ for small-$z$. For the asymptotic domain one has to
consider other methods. At present one can resort either to the Regge
phenomenology \cite{Regge}, or to the perturbative QCD analysis
\cite{For95,SMX}. While the former has purely phenomenological character, the
latter is still subject to an ongoing research and hence to some controversy
\cite{Con}.

Consider now the problem of construction of the quark momentum distribution
from the available theoretical input. For definiteness we shall consider only
the u-quark distribution, postponing the detailed discussion of the d-quark
distribution to a more detailed publication \cite{Inprep}.  We choose the
MRS(A) parametrization \cite{MRS} to represent the data.  Our goal is to
develop an approximation which allows to reproduce, with a reasonable accuracy,
the MRS(A) u-quark distribution at $\mu^2 = 4$ GeV$^2$ using modest amount of
theoretical information.
    
As the first step we consider the Ioffe-time distribution.  As discussed above,
it is natural to consider the small-$z$ and large-$z$ regimes separately. The
information necessary to determine $Q(z)$ in the the first region is contained
in a few lowest moments of the momentum distribution. We assume that they are
computable on the lattice, but of course the smaller is the required number of
moments the better. We shall assume that lattice QCD is capable of computing
these moments with a high accuracy and calculate their numerical values using
the MRS(A) parametrization. In the large-$z$ region we assume that the
standard, naive Regge argument is valid, namely that $q(u) \sim u^{-1}$ at
small u, which corresponds to $Q(z) \sim$ const. at large $z$. If the
normalization point $\mu^2$ is relatively low, it may not be a bad
approximation and, as it is shown below, it will influence only the small-$u$
behavior. Now, we have to match the small-$z$ behavior, given by an almost
straight line with the slope given by the first moment of $q(u)$ with the
asymptotic behavior given by the line $y$ = const., and for that we need
additional information about the behavior of $Q(z)$ in the transition region.
In the first approximation one can argue that the transition region occurs when
$z$ reaches the value which corresponds to the confinement radius $\sim 1$ fm
in the nucleon rest frame, or $z \approx 5$. Assuming that the transition is
infinitely sharp we have obtained the approximate Ioffe-time distribution which
is depicted by the dot-dashed line on figure 4. After transformation to
longitudinal momentum space it results in the approximation to $u q(u)$ denoted
by the dot-dashed line on figure 5. Note that this approximation, while
certainly not satisfactory, relies only on the momentum fraction carried by
quarks, which is nowadays computable in lattice QCD, and arguments about the
Regge behavior and confinement radius.

It is clear that in the next step one has to take into account more accurately
the {\em magnitude} of $Q(z)$ in the asymptotic domain. This is crucial, see
figure 4, for a successful reconstruction of the shape of the experimental
distribution. Two numbers are required to predict gross features of u-quark
distribution function. The first one is the u-quark momentum fraction, the
second is the large-$z$ magnitude of its Ioffe-time distribution. It can be
nicely demonstrated by matching the small-$z$ behavior given by the first
moment with the correct large-$z$ magnitude. The result is denoted by the
dotted line in figure 5. When the negative large-$u$ tail, which contradicts
the positivity requirement of a probability distribution, is neglected, the
approximation is quite satisfactory and probably much better than what one can
hope to obtain from many phenomenological models. While the momentum fraction
is computable on the lattice, we are not aware about any analysis which could
provide us with reliable information about the large-$z$ magnitude of $Q(z)$.
To proceed still further we have chosen to approximate $Q(z)$ more accurately
in the transition region using information about the higher moments of the
parton distribution. It can be done very efficiently using Pad{\'e}
approximation technique. Note that the shape of the u-quark Ioffe-time
distribution exhibits a maximum in the transition region which is a natural
candidate for a point where the curves determined by the small-$z$ and the
asymptotic behavior should match each other.  The improved description of
$Q(z)$ in the transition region based on the [3,2] Pad{\'e} approximation
results in the dashed curve on figure 4.  Note that it has been obtained using
the information about the first three non-vanishing moments and the Regge
argument about flat large-$z$ behavior.  The corresponding approximation to the
longitudinal momentum distribution $u q(u)$ is denoted by a dashed line in
figure 5. The agreement with the exact result, solid line, is satisfactory,
except for the small-$u$ region, where the true behavior is markedly different
from the assumed naive Regge asymptotics.

Calculation of the first three moments of the C = + 1 quark longitudinal
momentum distribution is equivalent to a lattice measurement of matrix elements
of operators built from a one gamma matrix and one, three, or five derivatives
between two quark fields. The first lattice QCD results on the first two
moments in the quenched approximation have been already reported \cite{Lat96}.
Unfortunately, due to technical difficulties it is not yet possible to measure
the third moment.

When the Ioffe-time momentum distribution has been obtained, it has to
be inverted according to the formula:
\begin{equation}
q(u) = \frac{2}{\pi} \int_0^\infty dz\, sin(uz) Q(z) \, ,
\label{inversion}
\end{equation}
to produce the experimentally measurable longitudinal momentum distribution.
Note that we are dealing with the positive C-parity combination of quark and
antiquark longitudinal momentum densities. Because $Q(z)$ is not square
integrable, the integral (\ref{inversion}) converges very slowly. A fast and
accurate calculational method is available if the the asymptotic behavior of
$Q(z)$ for large $z$ is known. Indeed, let us assume that $Q(z) \sim C z^\alpha
+ D$ when $z \to \infty$. Then, subtracting the asymptotic behavior of $Q(z)$
from the integrand in (\ref{inversion}) one easily obtains
\begin{eqnarray}
q(u) & = & \frac{2}{\pi} \int_0^\infty dz\, sin(uz) 
\left[ Q(z) - C z^\alpha - D \right] \nonumber \\
& + & \frac{2}{\pi} C \cos (\frac{\pi \alpha}{2}) 
\frac{\Gamma(1 + \alpha)} {u^{1+\alpha}} + 
\frac{2}{\pi} D \frac{1}{u} \, .
\label{inverse1}
\end{eqnarray}
Now the integral converges fastly and can be easily computed.
We have used this method to invert consecutive approximations
of Ioffe-time distributions discussed above.

Having in mind the problem of determination of the input distributions for
parametrizations of parton densities, we have compared the present method with
a fit to the first three moments which assumes some simple functional form. In
the full analogy with the actual shape of the input distribution to the MRS(A)
parametrization we have tested the functional form
\begin{equation}
u q (u) = A (1 - u)^b + C u ^d (1-u)^b \, .
\label{test}
\end{equation}
Equation (\ref{test}) as it stands depends on four parameters $A$, $b$, $C$,
and $d$. As it is not possible to determine all of them from the information
about three moments we decided to fix $d$ and to find three remaining
parameters. The best agreement with the original distribution is found for $d
\sim 1.5$. Because the MRS(A) parametrization at $\mu^2 = 4$ GeV$^2$ follows at
large $u$ exactly the shape (\ref{test}), the agreement with the approximate
form is perfect in this domain. The Regge assumption about small-$u$ behavior
leads to a discrepancy in this region. Note however that the method based on
the consideration of Ioffe-time distribution allows us actually to {\em derive}
the shape (\ref{test}) from first principles augmented by a Regge argument
about the large-$z$/small-$u$ behavior. Hence we have been able to show that
the input distribution can be strongly constrained by the information stemming
from non-perturbative QCD without having to rely on purely phenomenological
models. On the other hand it is tempting to conclude from our discussion that
traditional quark models of hadronic structure should attempt to compute first
few moments of parton distributions rather than their full Bjorken $x$
dependence.

Approximate determination of the d-quark distribution can be done using
essentially the same technique \cite{Inprep}. As the MRS(A) parametrization in
the transition region shows no maximum, there is no obvious choice of the
matching point between small-$z$ and large-$z$ behavior. As a consequence, a
reasonable approximation requires knowledge of the first {\em four} moments
i.e., one more than in the u-quark case. This problem can be circumvented in
the case of the CTEQ3 parametrization which is almost flat in the asymptotic
region and therefore fits exactly into our scheme.

Can the accuracy be further improved to reach, for example, the level of
techniques developed in \cite{Par79,Fur82,Kat94}? The answer is yes, despite of
the fact that the question has, from our point of view, purely academical
character. We have found \cite{Inprep} that Pad{\'e} approximation taking into
account the lowest six moments of the momentum distribution allows to
reconstruct $Q(z)$ reliably in the $z$ region which extents so far away from
the origin that the naive Regge argument is not more necessary and the true
asymptotic behavior can be found by inspection. The resulting approximations of
Ioffe-time and longitudinal momentum distributions are almost perfect.  We
argue, however, that such procedure, although mathematically consistent, is not
sensible from the physical point of view. Calculation of the sixth moment
requires e.g., a lattice measurement of the matrix element of a operator with
twelve Lorentz indices i.e., one gamma matrix and eleven derivatives.  Note
that on the lattice a derivative is replaced by a finite difference. Thus,
because of the size of such an operator and because of its mixing with
operators with lower dimension which occurs in the discretized lattice
formulation \cite{Mix96}, it would be prohibitively difficult to perform such a measurement
using the present technology.  In such situation it is natural to suggest that
one should develop a different technique to understand the shape and the
magnitude of a Ioffe-time distribution at large $z$. In this domain the
Compton scattering amplitude in the target rest frame is
dominated by the photon splitting into a quark-antiquark pair, which
subsequently scatters off the background color field of
the target \cite{Duc92}.  While it seems that the resulting shape of $Q(z)$ can
be understood within perturbative QCD \cite{SMX}, the magnitude is certainly a
problem of a non-perturbative character \cite{Bal95}. \\
\vskip 1 cm

{\bf Acknowledgments} We are grateful to R. Horsley for helpful discussions
about the present status of lattice QCD calculations and for critical reading
of the manuscript. This work was supported
in part by BMBF and by KBN grant 2~P03B~065~10.

\vfill\eject
\section*{Figure captions}
\begin{description}
  
\item[Fig.~1] $Q_u(z)$, the u-quark Ioffe-time distribution at the scale $\mu^2
  = 4$ GeV$^2$. The solid line denotes MRS(A) \cite{MRS}, the dashed line
  denotes CTEQ3 \cite{CTEQ}, and the dotted line denotes the
  Gl{\"u}ck,Reya,Vogt \cite{GRV94} parametrizations.

\item[Fig.~2] $Q_d(z)$, the d-quark Ioffe-time distribution at the scale $\mu^2
  = 4$ GeV$^2$. The solid line denotes MRS(A) \cite{MRS}, the dashed line
  denotes CTEQ3 \cite{CTEQ}, and the dotted line denotes the
  Gl{\"u}ck,Reya,Vogt \cite{GRV94} parametrizations.

\item[Fig.~3] Convergence of the Taylor expansion of $Q_u(z)$. The line
  labeled Tn denotes an approximation which requires the knowledge of the
  first n nonvanishing derivatives at the origin, or equivalently the moments
  of quark longitudinal momentum distribution.

\item[Fig.~4] Consecutive approximations of $Q_u(z)$, solid line labeled
  MRS(A).  The first approximation, dot-dashed line labeled T1, assumes a
  sharp transition between small-$z$ and asymptotic regimes at the value $z =
  5$, which corresponds to the confinement radius $\sim 1$ fm.  The next
  approximation, dashed line labeled P[3,2], takes into account the behavior
  of $Q(u)$ in the transition region more accurately by employing a Pad{\'e}
  approximation based on the first three moments.
  
\item[Fig.~5] Consecutive approximations to the u-quark momentum distribution
  $u q(u)$, solid line labeled MRS(A).  The dot-dashed line, labeled T1, is
  based on the approximation which relies on the value of the first moment of
  $q(u)$, the Regge behavior at small-$u$, and the magnitude of the
  confinement radius. The dotted line, labeled T1$^\prime$ results from
  matching of small-$z$ behavior given by the first moment with the correct
  large-$z$ magnitude. The dashed line, labeled P[3,2], requires the values of
  the first three moments of $q(u)$.

\end{description}

\clearpage

\epsfbox{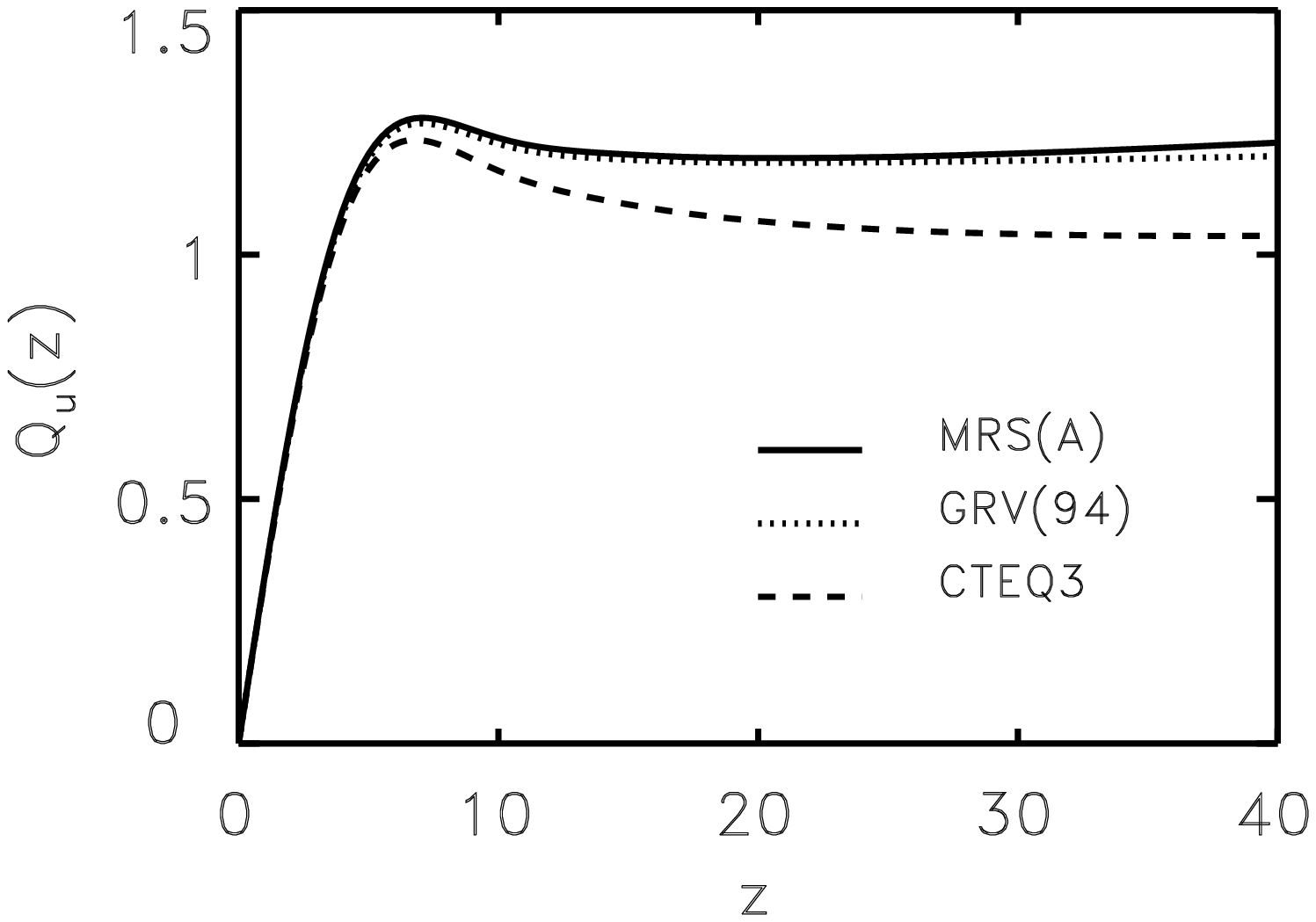}
\begin{description}
  
\item[Fig.~1] $Q_u(z)$, the u-quark Ioffe-time distribution at the scale $\mu^2
  = 4$ GeV$^2$. The solid line denotes MRS(A) \cite{MRS}, the dashed line
  denotes CTEQ3 \cite{CTEQ}, and the dotted line denotes the
  Gl{\"u}ck,Reya,Vogt \cite{GRV94} parametrizations.

\end{description}

\epsfbox{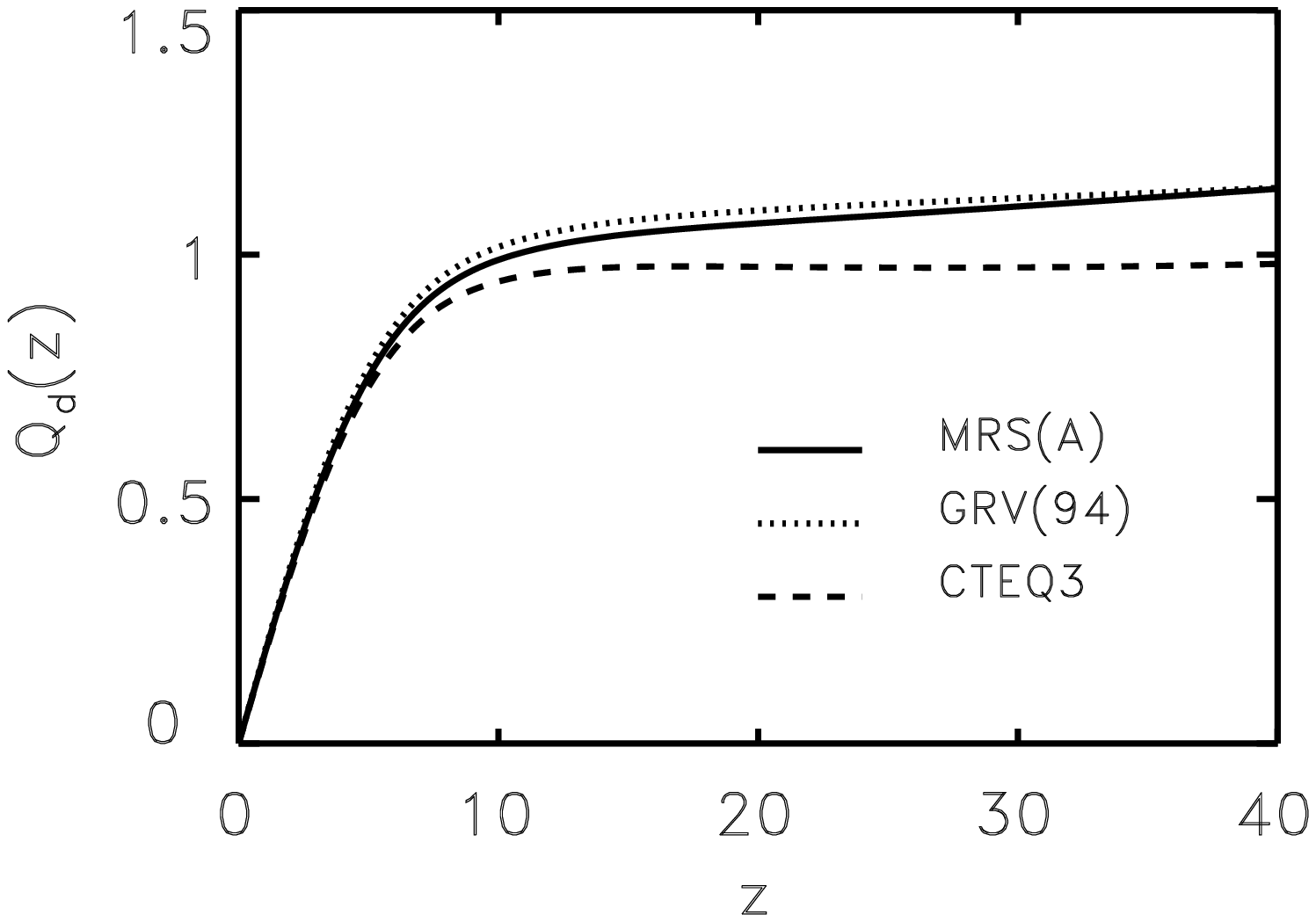}
\begin{description}
  
\item[Fig.~2] $Q_d(z)$, the d-quark Ioffe-time distribution at the scale $\mu^2
  = 4$ GeV$^2$. The solid line denotes MRS(A) \cite{MRS}, the dashed line
  denotes CTEQ3 \cite{CTEQ}, and the dotted line denotes the
  Gl{\"u}ck,Reya,Vogt \cite{GRV94} parametrizations.

\end{description}

\epsfbox{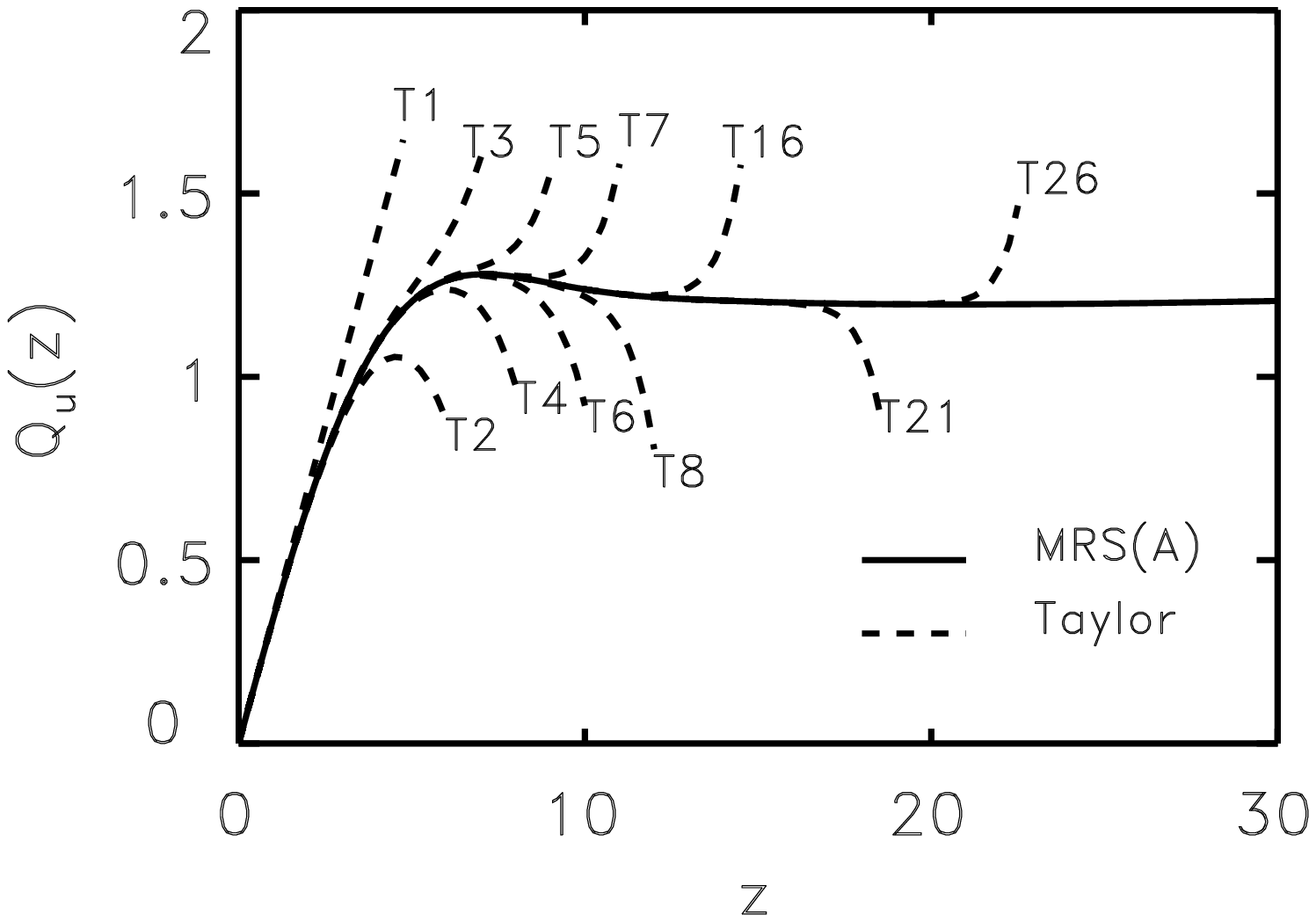}
\begin{description}
  
\item[Fig.~3] Convergence of the Taylor expansion of $Q_u(z)$. The line
  labeled Tn denotes an approximation which requires the knowledge of the
  first n non-vanishing derivatives at the origin, or equivalently the moments
  of quark longitudinal momentum distribution.

\end{description}

\epsfbox{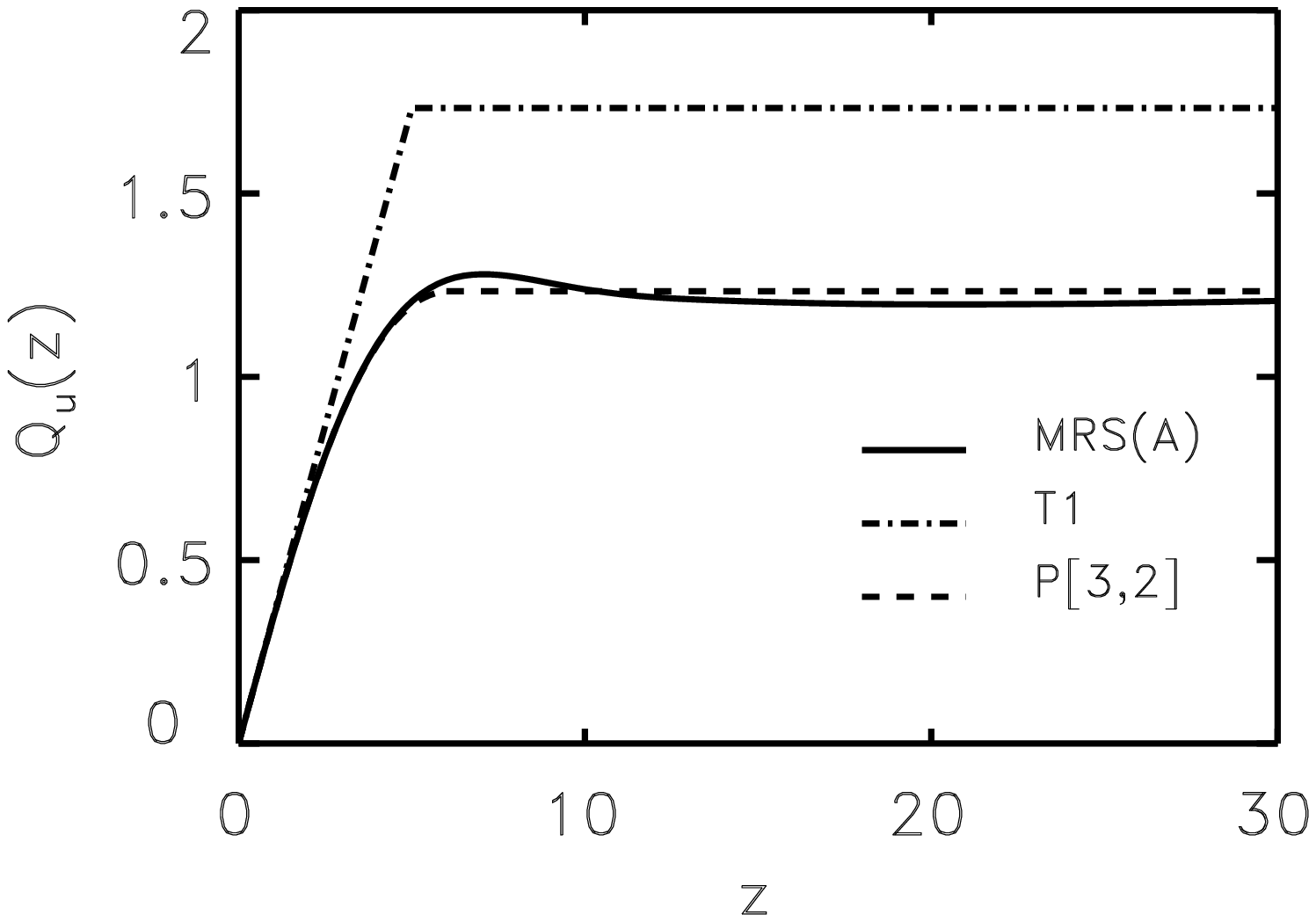}
\begin{description}
  
\item[Fig.~4] Consecutive approximations of $Q_u(z)$, solid line labeled
  MRS(A).  The first approximation, dot-dashed line labeled T1, assumes a
  sharp transition between small-$z$ and asymptotic regimes at the value $z =
  5$, which corresponds to the confinement radius $\sim 1$ fm.  The next
  approximation, dashed line labeled P[3,2], takes into account the behavior
  of $Q(u)$ in the transition region more accurately by employing a Pad{\'e}
  approximation based on the first three moments.

\end{description}

\epsfbox{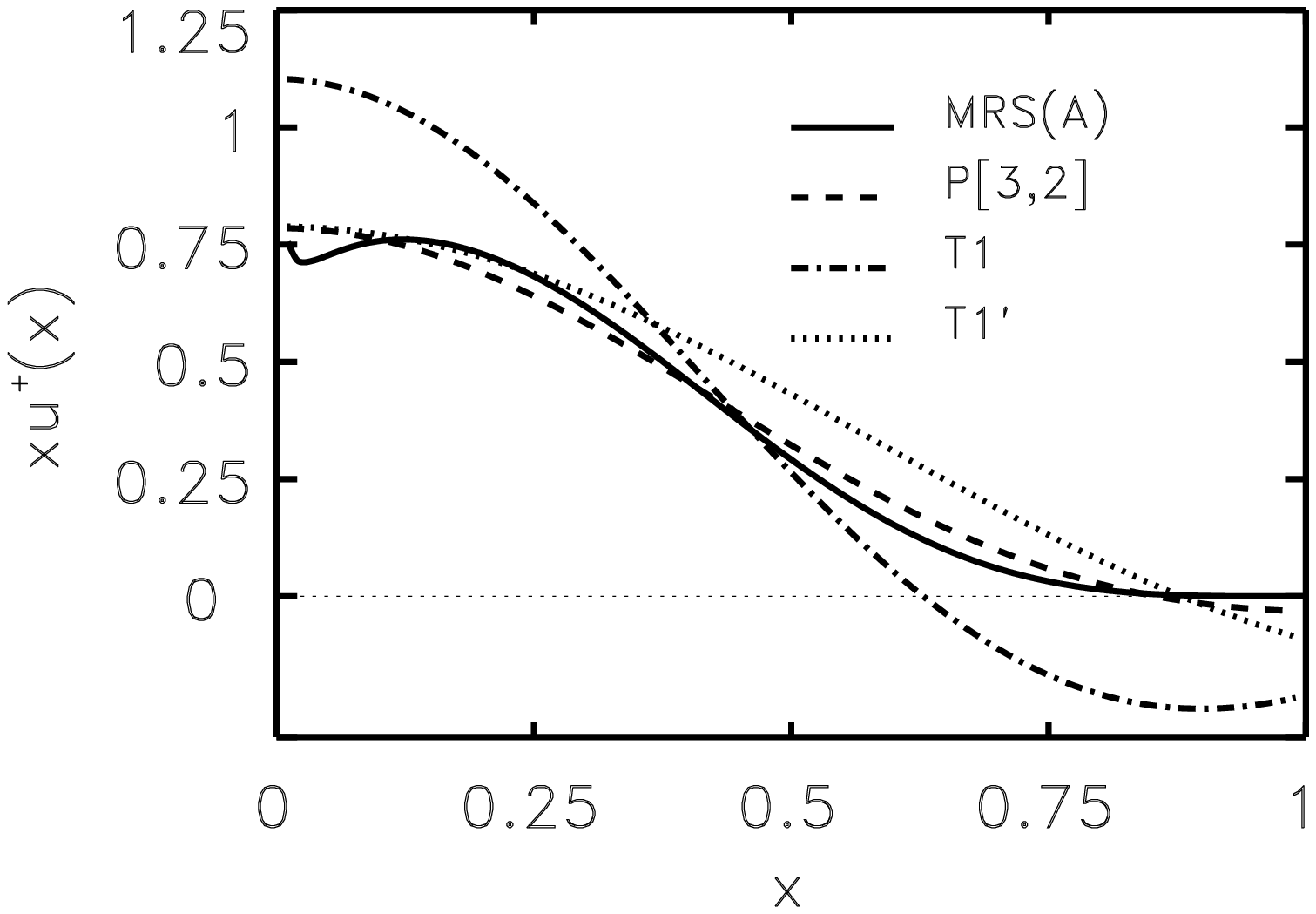}
\begin{description}
  
\item[Fig.~5] Consecutive approximations to the u-quark momentum distribution
  $u q(u)$, solid line labeled MRS(A).  The dot-dashed line, labeled T1, is
  based on the approximation which relies on the value of the first moment of
  $q(u)$, the Regge behavior at small-$u$, and the magnitude of the confinement
  radius. The dotted line, labeled T1$^\prime$ results from matching of the
  small-$z$ behavior given by the first moment with the correct large-$z$
  magnitude. The dashed line, labeled P[3,2], requires the values of the first
  three moments of $q(u)$. Please note that the corresponding dotted curve in
  our published version of the paper is wrong due to an error in our computer 
  program.

\end{description}

\end{document}